\documentclass[useAMS,usenatbib]{mnras}

\usepackage{bm}
\usepackage{natbib}
\usepackage{graphicx}
\usepackage{amssymb, amsmath}
\usepackage{ulem,color}

\newcommand{\acdel}[1]{}

\title[Crucial role of neutron diffusion]{Crucial role of neutron diffusion in the crust of accreting neutron stars}
\author[A.I.\ Chugunov, N.N.\ Shchechilin]
{A.I.\ Chugunov and N.N.\ Shchechilin
\thanks{nicknicklas@mail.ru}\\
$^1$Ioffe Institute, Politekhnicheskaya 26, 194021 Saint Petersburg, Russia}

\begin{document}
	
	\date{Accepted 2020 March 25. Received 2020 March 23;
		in original form 2020 February 28}
	
	\pagerange{\pageref{firstpage}--\pageref{lastpage}}
	\pubyear{2018}
	
	\maketitle
	
	\label{firstpage}

\begin{abstract}
Observed temperatures of transiently accreting neutron stars in the quiescent state are generally believed to be supported by deep crustal heating, associated with non-equilibrium exothermic reactions in the crust. Traditionally, these reactions are studied by considering nuclear evolution governed by compression of the accreted matter. Here we show that this approach has a basic weakness, that is that in some regions of the inner crust the conservative forces, applied for matter components (nuclei and neutrons), are not in mechanical equilibrium. In principle the force balance can be restored by dissipative forces, however the required diffusion fluxes are of the same order as total baryon flux at Eddington accretion. We argue that redistribution of neutrons in the inner crust should be involved in realistic model of accreted crust. 
\end{abstract}


\section{Introduction}

The inner regions of neutron stars (NSs) are composed of matter with density exceeding the density in atomic nuclei and interpretation of NS observations provides a unique opportunity to check theoretical models of matter at such extreme conditions (e.g., \citealt*{hpy07}; \citealt{CompStarBook18}). NSs are observed in different environments, which mimics different ``experiments'' performed by nature with these objects. In particular, some NSs are found in so-called low mass X-ray binaries, i.e. they are located in binary systems, where the companion star fills the Roche lobe and transfers matter to NS. In some cases, the matter falls to NS surface not permanently, but in a transient way. 
In quiescent periods one is able to observe the thermal emission from NS surface and, thus, to measure its surface temperature (see e.g.\ \citealt*{pcc19} for the recent compilation of observational results).
These observations support the deep crustal heating paradigm by \cite*{bbr98}:  accretion heats up the NS crust by non-equilibrium exothermic nuclear reactions, which occurs in previously accreted material while it is being buried by freshly accreted matter.

Studies of respective non-equilibrium reactions and composition of accreted NSs were started by \cite{Sato79}, then continued in a series of papers by \cite{HZ90,HZ90b,HZ03,HZ08,Fantina_ea18} within one-component model.
Multicomponent models were considered, e.g.,  by \cite{Gupta_ea07,Steiner12,lau_ea18,SC19_JPCS,SC19_MNRAS}.
However, all previous works follow the traditional approach --  they suppose that all changes in composition are associated with nuclear reactions inside a portion of accreted matter and that these reactions are induced only by compression.
\cite{Steiner12} pointed that this approach might be inconsistent: the one-component approximation leads to the  jumps in the number density of free neutrons in the inner crust, which are not totally smoothed out in multi-component models. In spite that the neutron diffusion in young NSs was considered in early works by \cite*{BKKC76,bkc79}, typically, the inconsistency was more or less ignored, perhaps, because this problem was assumed to be a local one, the one, which does not affect global structure of the crust significantly. 

In this letter, we show this to be incorrect.
For simplicity and physical transparency, here we assume that free  neutrons in inner crust are nonsuperfluid (see footnote \ref{footnote_SF}; the neutron superfluidity is taken into account in subsequent work \citealt{GC20_DiffEq}).
 Namely, we consider the crust structure within traditional approach (layers of compressed accreted matter, one located on top of another to be in net hydrostatic equilibrium) and demonstrate that the diffusion equilibrium for the components (atomic nuclei, electrons, and neutrons) is strongly violated in some finite regions of the inner crust. In particular, in some layers, all conservative forces (electrostatic and gravitational) applied to nuclei are directed downwards and
only dissipative forces can prevent nuclei from falling down and restore the force balance.
We argue that this indicates significant diffusion fluxes of neutrons, which occur in the inner crust of accreting NSs and lead to redistribution of nucleons inside it.
Obviously, this process should affect the composition. As a result, the nuclear processes, equation of state (EOS), and deep crustal heating profile should also be modified.

\section{Diffusion equilibrium in the  crust}
\subsection{Physics input} \label{Sec_PhysInput}

Following \cite{SC19_MNRAS}, we apply several tabulated nuclear models, marked as 'FRDM95+DZ31', 'FRDM12+DZ31', 'HFB21+DZ31', and 'DZ31'.
Namely, the atomic mass evaluation 2016 (AME16; \citealt{ame16}) is used
for all experimentally known masses,
while the masses of other nuclei were taken from two versions of finite-range droplet macroscopic model (FRDM92, \citealt{FRDM95} and FRDM12, \citealt{FRDM12}),  Hartree-Fock-Bogoliubov (HFB) calculations of  the energy density functional  BSk21 (HFB21, \citealt{HFB21}), and DZ31 -- 31-parameter model by \cite*{DZ95,Anatomy_DZ_10}, respectively.%
\footnote{
	The table for AME16 and code for DZ31 model were downloaded from \url{https://www-nds.iaea.org/amdc/}. The table for FRDM92 model was downloaded from \url{http://t2.lanl.gov/nis/molleretal/publications/ADNDT-59-1995-185-files.html}.
	HFB21 data downloaded from \url{http://www-astro.ulb.ac.be/bruslib/nucdata/hfb21-dat}, \cite{bruslib}.}
DZ31 mass model was also applied beyond available FRDM95, FRDM12, and HFB21 tables.
We also use the compressible liquid drop model by \cite{MB77}  (MB in what follows).

The energy density can be written as
\begin{equation}
\epsilon=\sum_\alpha n_\alpha m_\alpha(Z_\alpha,A_\alpha)
+\epsilon_\mathrm{e}(n_\mathrm{e})
+\epsilon_\mathrm{n}(n_\mathrm{n})
+\epsilon_\mathrm{L},
\end{equation}
where $\epsilon_\mathrm{e}(n_\mathrm{e})$ and $\epsilon_\mathrm{n}(n_\mathrm{n})$ are energy density of electrons (assumed to be free and fully degenerate) and neutrons respectively.  $m_\alpha(Z_\alpha,A_\alpha)$ is the mass of nucleus with charge number $Z_\alpha$ and mass number $A_\alpha$. Within MB model $m_\alpha$ depends also on the density of free neutrons $n_\mathrm{n}$, but this dependence is weak for considered pressure region, and  neglected below as well as the volume, occupied by nuclei, when calculating the energy density associated with free neutrons. So-called lattice term $\epsilon_\mathrm{L}$, associated with Coulomb interaction of atomic nuclei with electron background and other nuclei, will be also neglected (except the section \ref{Sec_GenNetwork}) because it is small in comparison with $\epsilon_\mathrm{e}$. 
These simplifications shorten the derivations significantly and do not affect the main results, because the  diffusion equilibrium equations are violated on the order of the leading term $\nabla \mu_\mathrm{e}$.

The pressure is given by 
%
$P=-\partial \epsilon V/\partial V
=P_\mathrm{e}(n_\mathrm{e})+P_\mathrm{n}(n_\mathrm{n})$. 
%
Here $P_\mathrm{e}(n_\mathrm{e})=-\epsilon_\mathrm{e}+n_\mathrm{e} \mu_\mathrm{e}$ 
and
$P_\mathrm{n}(n_\mathrm{n})=-\epsilon_\mathrm{n}+n_\mathrm{n} \mu_\mathrm{n}$ 
are pressures of electrons and neutrons respectively;
$\mu_\mathrm{e}=\partial \epsilon_\mathrm{e}/\partial n_\mathrm{e}$ 
and
$\mu_\mathrm{n}=\partial \epsilon_\mathrm{n}/\partial n_\mathrm{n}$ 
are respective chemical potentials, which include the rest masses. 
The chemical potential on nuclei $\mu_{\alpha}=m_\alpha(Z_\alpha,A_\alpha)$.

\subsection{Force balance equations} \label{Sec_DifEq}
Let us write down the force balance equations for the components of the inner crust: electrons, nuclei, and free neutrons, assumed to be nonsuperfluid for simplicity%
\footnote{As discussed in \cite{kg18} (around the footnote 2 there), the dissipation effects for superfluid neutrons at finite temperature can not be described in form of simple force balance equations. We take an opportunity to thank M.E.~Gusakov for pointing this feature to us and refer reader to subsequent paper \cite{GC20_DiffEq}, there neutron superfluidity is taken into account consistently. \label{footnote_SF}
}
 (see \citealt{by13} for the case of outer crust)
\begin{eqnarray}
e\nabla \phi +m_{\mathrm e}^\ast{\bm{g}} -{{\nabla}}\mu_{\mathrm e}&=& \bm f_\mathrm{ ei}  
+ \bm f_\mathrm {en} , \label{de}\\
-eZ\nabla \phi+m_\mathrm{i} {\bm{g}}-\sum_{\alpha}\frac{n_\alpha}{n_\mathrm{i}}{{\nabla}}\mu_{\alpha}&=&
\bm f_\mathrm{ie}+\bm f_\mathrm{in}, \label{di}\\
m_n {\bf{g}}-{\bf{\nabla}}\mu_n&=&\bm f_\mathrm{ni}  
+ \bm f_{\mathrm {ne}}. \label{dn} 
\end{eqnarray}
Here equation (\ref{di}) presents the sum over force balance equations for all nuclei types $\alpha$, $Z$ is an average charge of nuclei, $\bm g$ is  gravitational acceleration; finally $m_j$, $\mu_j$, and $\bm f_{jk}$ are mass, chemical potential, and dissipative force from particles of type $j$ normalized  to one particle of type $k$ ($j,k=\mathrm{e,\ i, \ \alpha,\ n}$; index $\mathrm i$  means averaging over nuclei types). 
Electrons are relativistic, thus equation (\ref{de}) contains effective mass $m_\mathrm e^\ast=\mu_\mathrm e$ (e.g.\, \citealt{Passamonti_ea17}), for other particles the difference between effective mass and the rest mass can be neglected.
Summing equations (\ref{de}-\ref{dn}), multiplied to the respective number densities $n_i$, we come to 
hydrostatic equilibrium equation 
%
$ \nabla P=\rho g$. 
%
Here we introduce the total mass density $\rho=\sum m_j n_j$.
We also take into account: (a) the quasi-neutrality condition ($n_\mathrm e=Z n_\mathrm i$), (b) the Gibbs-Duhem relation ($\mathrm d P=n_\mathrm e  \mathrm d\mu_\mathrm e+ n_\mathrm i  \mathrm d\mu_\mathrm i+n_\mathrm n \mathrm  d\mu_\mathrm n$), and  (c) the third Newtonian low $n_i \bm f_{ik}=-n_k \bm f_{ki}$.

Expressing $\bm g$ from hydrostatic equilibrium equation 
and substituting it into sum of equations (\ref{de}) and (\ref{di}), multiplied to respective number densities, we arrive to 
\begin{equation}
\sum_{\alpha} n_\alpha{\nabla}\mu_{\alpha} +n_\mathrm e\nabla \mu_\mathrm e =
 \frac{n_\mathrm i m_\mathrm i + n_\mathrm e m^\ast_\mathrm e}{\rho} \nabla P
 -n_\mathrm{i} f_\mathrm{in}-n_\mathrm{e} f_\mathrm{en}.
 \label{dP_equl}
\end{equation}
With simplifications of section \ref{Sec_PhysInput}, $n_\alpha{\nabla}\mu_{\alpha}=0$. Neglecting  $n_\mathrm e m^\ast_\mathrm e$ in comparison with $n_\mathrm i m_\mathrm i$ we get
\begin{equation}
\nabla \mu_\mathrm e =
\frac{m_\mathrm i}{Z\,\rho} \nabla P
-\frac{1}{Z} f_\mathrm{in}-f_\mathrm{en}, \label{gradmu_gen}
\end{equation}
where the quasi-neutrality condition
was applied. 
The forces $\bm f_{jk}$ in equations (\ref{de}--\ref{dn}) are dissipative, associated with 
friction, which occurs, if there is a flow of (nonsuperfluid) particles of type $j$ with respect to particles $k$. So, if there are no such flows (i.e. crust is in diffusion equilibrium), these forces should be vanishing.
In this case the derivative $\partial \mu_{\mathrm e}/\partial P$ along the equation of state should be equal to
\begin{equation}
\left .\frac{\partial \mu_{\mathrm e}}{\partial P}\right |_\mathrm{NoDiff}=\frac{ m_\mathrm i}{Z \rho}\approx \frac{A m_\mathrm U}{Z \rho}. \label{NoDiffGrad}
\end{equation}
In the last equality we estimate $m_\mathrm i\approx A m_\mathrm{U}$, where $A$ is the averaged atomic number and $m_\mathrm{U}$ is atomic mass unit.

\subsection{Violation of diffusion equilibrium  in traditional approach}

The one-component models of accreted NS crusts assume that matter is composed of one type of atomic nuclei at each layer, which is known to cause the jumps of neutron number density. As pointed by \cite{Steiner12}, it is a clear evidence that one-component models are inconsistent with respect to thermodynamics (see also footnote \ref{footnote_OneComp}). Furthermore, even if the outer layers of the crust are one-component, the inner layers are driven to be multi-component by neutron emission/absorption reactions (e.g., \citealt{Steiner12,lau_ea18,SC19_JPCS,SC19_MNRAS}). Hence, in this section, we consider multi-component models.
We follow the traditional approach and trace compositional changes in the accreted material associated with nuclear transformations driven by compression resulting from increasing pressure. We demonstrate that this leads to a violation of the diffusion equilibrium condition in the crust.

\subsubsection{Simplified reaction network}
To begin, let us illustrate the absence of diffusion equilibrium at the accreted crust by using simplified nuclear reaction network described in \cite{SC19_MNRAS}. Within this approach, following \cite{Steiner12}, we consider only energetically favorable reactions and describe kinetics of these reactions in a step-wise manner -- at each step the reaction proceeds for a small fraction of nuclei. If several reactions are allowed, the proceeding reaction is selected according to the priority rules, based on typical reaction timescales. If none of the reactions is allowed, the pressure is increased until it reaches the threshold value to allow the next nuclear transformation (see \citealt{SC19_MNRAS} for details and explicit formulation of the priority rules).

\begin{figure}
	\includegraphics[width=\columnwidth]{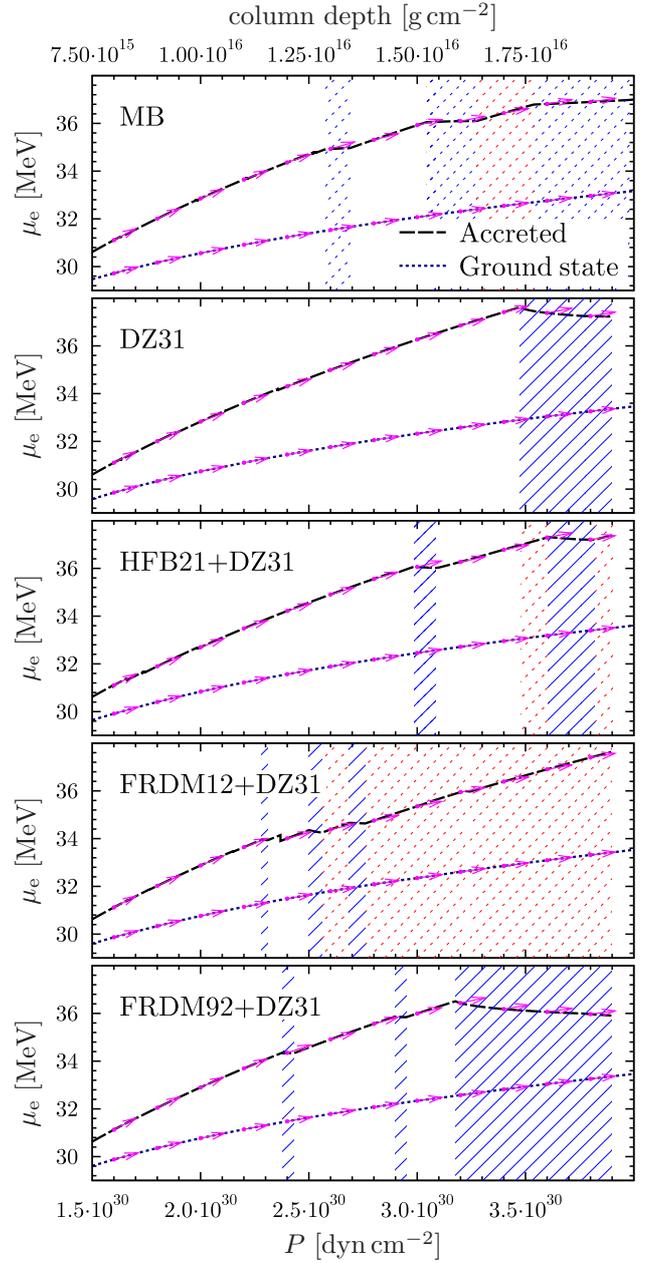}
	\caption{The profiles of electron chemical potential $\mu_e$ as function of $P$ for several nuclear physical models. Long dashes represent accreted crust, considered within multi-component version of traditional approach, dotted lines are for ground state composition.
		At several points, indicated by dots, the derivative $\left.\frac{\partial \mu_\mathrm{e}}{\partial P}\right|_\mathrm{NoDiff}$ is shown
		 (see the text for details). 
		 The regions of strong violation of diffusion equilibrium are hatched.
		The column density is shown at the upper horizontal axis (fiducial $g=2\times 10^{14}$~cm\,s$^{-2}$ is assumed).
	}
	\label{Fig_mue_profile}
\end{figure}

The composition of the crust for tabulated mass models was studied in \cite{SC19_MNRAS} and here, for illustration, we apply the results of `unmixed approach'
(see \citealt{SC19_MNRAS} for details).
The resulting profiles of  $\mu_\mathrm{e}$
as a function of pressure $P$ are shown by long dashes in figure \ref{Fig_mue_profile}.
The upper panel presents the $\mu_\mathrm{e}(P)$ profile obtained within the simplified reaction network for compressible liquid drop model by \cite{MB77}.%
\footnote{For MB model the accreted crust was recalculated according to priority rules formulated by  \cite{SC19_MNRAS}. In \cite{SC19_JPCS} we apply a bit different priority rules, which leads to  minor differences in composition of the accreted crust, but do not affect the features considered here.
}

For all tabulated models there are regions, where  $\mu_\mathrm{e}$ decreases with the increase of the pressure (shaded by solid lines).
The pressure is continuously increasing with depth, thus $\nabla \mu_\mathrm{e}$ is directed outwards in these regions. According to (\ref{de}), $\nabla \phi$ should be also directed outwards, but in this case all terms in the left-hand-side of (\ref{di}) are  directed downwards and can  be balanced only by the dissipative forces. Strong violation of the diffusion equilibrium condition takes place also in other parts of the accreted crust, which are shown by shading: the red-shaded regions correspond to $\frac{\partial \mu_{\mathrm e}}{\partial P}>1.5\, \left .\frac{\partial \mu_{\mathrm e}}{\partial P}\right |_\mathrm{NoDiff}$, while in the blue-shaded regions $\frac{\partial \mu_{\mathrm e}}{\partial P}<0.5\, \left .\frac{\partial \mu_{\mathrm e}}{\partial P}\right |_\mathrm{NoDiff}$. Only layers thicker than $1$~m for fiducial $g=2\times 10^{14}$~cm\,s$^{-2}$ are shaded.
For
visual demonstration,  at several values of pressure we display the equilibrium derivative $\left .\frac{\partial \mu_{\mathrm e}}{\partial P}\right |_\mathrm{NoDiff}$
by arrows. 
The actual behavior of $\mu_\mathrm{e}(P)$ clearly disagrees with $\left .\frac{\partial \mu_{\mathrm e}}{\partial P}\right |_\mathrm{NoDiff}$ in shaded regions.%
\footnote{For traditional one component models $\mu_\mathrm{e}(P)$ disagrees with $\left .\frac{\partial \mu_{\mathrm e}}{\partial P}\right |_\mathrm{NoDiff}$ inside shells between nuclear reaction layers, revealing that these shells are out of diffusion equilibrium.
\label{footnote_OneComp}
}

To demonstrate that this disagreement is not associated with some fault in our treatment of diffusion equilibrium, we plot the $\mu_\mathrm{e}(P)$ profiles for ground state composition (dotted lines), using the same nuclear model and indicate respective $\left .\frac{\partial \mu_{\mathrm e}}{\partial P}\right |_\mathrm{NoDiff}$ by arrows.%
\footnote{Value $\left .\frac{\partial \mu_{\mathrm e}}{\partial P}\right |_\mathrm{NoDiff}$, given by (\ref{NoDiffGrad}), depends on the composition, thus it generally differ for accreted and ground state crust, even if it is calculated at the same pressure.} 
The ground state crust is indeed in diffusive equilibrium,
which is the predictable result -- the ground state crust corresponds to the minimal energy, thus the transfer of particles inside the crust cannot decrease energy. Consequently, the diffusion flows can not arise because they are not energetically favorable.

\subsubsection{General reaction network} \label{Sec_GenNetwork}
In this section, we argue that the absence of diffusion equilibrium is not a specific feature of our simplified reaction network \cite{SC19_JPCS,SC19_MNRAS}, but it is a general feature of the traditional approach. First of all, in \cite{SC19_MNRAS} we demonstrate that the composition of inner crust obtained within detailed reaction network by \cite{lau_ea18} for initial $^{56}$Fe composition is well reproduced by our approach, thus the $\mu_{\mathrm e}$ profile, corresponding to results of that work, based on FRDM95+DZ31 model, should be close to respective profile in figure \ref{Fig_mue_profile}, being not in diffusion equilibrium. In particular,
\cite{lau_ea18} reports increase of $\mu_{\mathrm e}$ only of $0.1$ MeV on course of density increase from $\rho=1.7\times 10^{12}$~g\,cm$^{-3}$  to $\rho=2.4\times 10^{12}$~g\,cm$^{-3}$, which should correspond to the region of weak dependence of $\mu_\mathrm{e}$ on $P$  in their calculations.

The second argument probably is too general to be very convincing as it is: the accreted crust is not in thermodynamical equilibrium, thus redistribution of nucleons inside crust can be energetically favorable. However, the traditional approach does not consider the diffusion equilibrium condition (\ref{NoDiffGrad}), which is an independent requirement and can be satisfied by traditional solution only just by chance.

To make this argument more specific, we rewrite equations (\ref{de}-\ref{dn}) and demonstrate that in diffusion equilibrium they are equal to two hydrostatic equations for noninteracting liquids -- electrons+nuclei ($\mathrm{ei}$) and neutrons.
Namely, let sum equations (\ref{de}) and (\ref{di}), multiplied by  respective number densities and multiply (\ref{dn}) to $n_\mathrm{n}$. It leads to
\begin{eqnarray}
 \nabla P_\mathrm{ei}&=&\rho_\mathrm{ei}\,\bm g, \label{HydStat_ei}\\
 \nabla P_\mathrm{n}&=&\rho_\mathrm{n}\bm g \label{HydStat_n}
\end{eqnarray}
where we introduce densities $\rho_\mathrm{ei}=n_\mathrm i m_\mathrm i + n_\mathrm e m^\ast_\mathrm e$,  $\rho_\mathrm{n}=n_\mathrm n m_\mathrm{n}$ and pressure $P_\mathrm{ei}=\partial \epsilon_\mathrm{ei} V/\partial V$. 
Note, decoupling of equations (\ref{HydStat_ei}) and (\ref{HydStat_n}) does not require neglection of $\epsilon_\mathrm{L}$ term in energy density.
However, if one takes into account dependence of nuclei masses on $n_\mathrm{n}$, it generally would lead to coupling of the above equations.
Nevertheless, this coupling is small in considered pressure region of MB model and exactly absent for models bases on atomic mass tables, like  
the reaction network in \cite{lau_ea18},  because nuclei masses assumed to be independent on $n_\mathrm{n}$ by construction.

Within the traditional approach, one follows the evolution of the accreted element on course of increasing of total pressure $P=P_\mathrm{ei}+P_\mathrm{n}$. In these models the electron captures can be accompanied by neutron emissions. The emitted neutrons stays free, being  not captured by other nuclei above some value of pressure, representing outer and inner crust boundary. On course of these reactions the number density of neutrons increases, increasing neutron pressure in accordance with reaction rate (see, e.g., figure 7 in \citealt{lau_ea18} for neutron mass fraction as a function of density for several initial compositions), but not necessary this profile agrees with equation (\ref{HydStat_n}). The most pronounced case of violation of (\ref{HydStat_n}) is associated with pycnonuclear reactions. Namely, within the multicomponent model, these reactions take place in some pressure region, being accompanied by beta captures and neutron emissions, which drive the daughter nucleus to the initial one. Consequently, the net result of the pycnonuclear reaction is the conversion of one atomic nucleus into neutrons (e.g.\ \citealt{lau_ea18}). 
The neutron number density is increased in the pycnonuclear region until all nuclei, participating in these reactions, are converted to neutrons (or increased neutron chemical potential prevent the disintegration of daughter nucleus to initial ones by neutron emissions). The associated increase of $ P_n$ gives the thickness of the pycnonuclear region, which is finite.

Pycnonuclear reactions can proceed directly (e.g. $^{40}$Mg$+^{40}$Mg) or being triggered by beta capture (e.g. beta-capture by $^{40}$Mg triggers set of electon captures and neutron emissions leading to $^{25}\mathrm{N}+^{40}$Mg reaction in \citealt{lau_ea18}).
Let us start with the second case. The rate of pycnonuclear reaction is determined by the rate of the trigger -- the electron capture. The latter rate should be of the same order as the compression rate because on the opposite case 
associated neutron emission should lead to a rapid increase of $P_\mathrm{n}$ and the total pressure. Thus, the electron chemical potential should be a bit below beta-capture threshold, being almost constant until the total burnout of these nuclei. This statement agrees with already cited results by \citealt{lau_ea18}: the $\mu_{\mathrm e}$ increases only of $0.1$ MeV on course of density increase from $\rho=1.7\times 10^{12}$~g\,cm$^{-3}$  to $\rho=2.4\times 10^{12}$~g\,cm$^{-3}$, where pycnonuclear reactions occur.
As far as, $P_\mathrm{ei}$ is mostly determined by electrons, it cannot increase strongly.
Consequently, equation (\ref{HydStat_ei}) should be violated.

If pycnonuclear reactions proceed directly, their rate depends exponentially  on $n_\mathrm{e}$ (e.g., \citealt{Yakovlev_ea06}; \citealt*{cdy07_NucFus}). Consequently,  $n_\mathrm{e}$ should not increase strongly in the pycnonuclear burning region to avoid too rapid burn out of nuclei. Thus, the arguments similar to the written above can be applied to support the inevitable violation of the diffusion equilibrium.

\section{Does the absence of diffusion equilibrium indicates the real problem of the traditional approach?}

Let us estimate the diffusion flows which should arise in shaded regions of figure \ref{Fig_mue_profile}. 
Following \cite{BKKC76,bkc79} we neglect electron-neutron friction (assume $f_\mathrm{ne}\ll f_\mathrm{ni}$) and estimate the neutron current with respect to nuclei as
\begin{equation}
J_n\approx \frac{n_\mathrm{n}}{n_\mathrm{i}\sigma_\mathrm{ni} v_\mathrm{n}}\frac{ f_\mathrm{ni}}{m_\mathrm{n}},
\end{equation}
where numerical coefficients of order of unity are omitted, $\sigma_\mathrm{ni}\approx 10^{-23} (1+A^{1/3})^2$~cm$^{2}$ is neutron-ion scattering cross section and $v_\mathrm{n}\approx 1.6 \times 10^9 (\rho_\mathrm{n,\,12})^{1/3}$~cm\,s$^{-1}$ is neutron velocity ($\rho_\mathrm{n,\,12}=m_\mathrm{n} n_\mathrm{n}/10^{12}\mbox{\,g\,cm}^{-3}$).
To estimate $f_\mathrm{ni}$, we apply equation (\ref{gradmu_gen}) and the third Newtonian law,  
which leads to
%
$f_\mathrm{ni}\approx A n_\mathrm{i} m_\mathrm{U} g/n_\mathrm{n}$
%
and finally
\begin{equation}
J_n\approx \frac{A } {\sigma_\mathrm{ni} v_\mathrm{n}} g
\approx 3\times 10^{28} g_{14} A_{70}^{1/3} \rho_\mathrm{n,\,12}^{-1/3}\,\mbox{cm}^{-2}\,\mbox{s}^{-1}.
\end{equation}
Here $g_{14}=g/10^{14}\,\mbox{cm\,s}^{-2}$ and $A_{70}=A/70$.

Let us compare $J_n$ with the total baryon flux $J_\mathrm{acc}$ associated with accretion. For the Eddington accretion rate $\dot m^\mathrm{E}=10^5$~g\,cm$^{-2}$\,s$^{-1}$, $J^\mathrm{E}_\mathrm{acc}=\dot m^\mathrm{E} /m_\mathrm{U}=6\times 10^{28}$~cm$^{-2}$\,s$^{-1}$ is of the same order as $J_n$, suggesting that neutron diffusion can redistribute baryons inside crust at the similar rate as accretion flow (for many NSs the  accretion rate, in fact, is  much slower, e.g., \citealt*{Done_ea07}). Furthermore, as far as neutron diffusion should also act in quiescent periods, it is more reliable to compare neutron diffusion flux with long time averaged accretion rate, which typically is at least two orders of magnitude lower than Eddington rate for transient sources (see e.g., \citealt{pcc19}).  In this case  the neutron diffusion predicted to redistribute nucleons even faster than accretion and thus should affect compositional evolution of the crust crucially.

\section{Summary}

In this letter, we demonstrate that the traditional approach for modeling accreted NS crust leads to strong violation of diffusion equilibrium condition in the crust.  In principle, non-superfluid neutrons, considered here for simplicity 
(see footnote \ref{footnote_SF}), can have enough strong diffusion flows to restore the mechanical equlibrium by friction, but the neutron flux  should be of the same order as net baryon flux corresponding to Eddington accretion and thus should affect crustal composition crucially.
The reason of this problem is an implicit assumption of the traditional approach, that all compositional changes are associated with  nuclear reactions, induced by compression of accreted matter and none of the nucleons can escape from compressing volume.
This assumption is clearly valid at the outer crust, where all nucleons are confined in nuclei, but does not work for the inner crust, where unbound neutrons can move between crustal layers. 
As a result, the nuclear evolution of the crust becomes a combined diffusion-nuclear burning problem, which is generally very complicated. 

Hopefully, as we argued here, the neutron diffusion in the crust 
typically 
redistributes neutrons faster than the accretion. As a result, one can consider the inner crust in the limit of fast diffusion, i.e. model nuclear processes enforcing diffusion equilibrium condition (the traditional approach corresponds to the opposite slow diffusion limit).

It should be noted, that here we consider only outer layers of the inner crust, where one can neglect the effects of unbound neutrons on nuclei masses. In the deeper layers, this approximation becomes invalid and more elaborated consideration is required. 
Such consideration is performed in subsequent paper \cite{GC20_DiffEq}, which also takes into account the neutron superfluidity effects and present a diffusion equlibrium model for fully accreted crust.
In particular, it demonstrates that almost all properties of diffusion equilibrium crust (equation of state, composition, heat release, etc.) are significantly different from the result of the traditional approach.

As a final remark, let us point that construction of diffusion equilibrium crust, in principle, can be important for the shallow heating problem -- the phenomenological powerful heating source of unknown nature, which is introduced to the models of crustal cooling to explain observational data (\citealt{mdkse18,Degenaar_ea19_AqlX1,Parikh_ea19_MXB1659}). This source is typically assumed to be localized in outer crust, thus it is rather unlikely that neutron diffusion provides a direct mechanism to explain this source. However, up-to-date constraints to its location are based on the results of the traditional approach and should be updated for a more realistic diffusion equilibrium solution.

\section*{Acknowledgements}
We are grateful to  M.E.~Gusakov, D.G.~Yakovlev, D.D.~Ofengeim,   K.P.~Levenfish, and P.S.~Shternin for useful discussions.  Work is supported by Russian Science Foundation (grant 19-12-00133).

\label{lastpage}

\end{document}